\documentclass[sigconf,balance=false]{acmart}
\usepackage{popets}

\setcopyright{popets}
\copyrightyear{2023}

\acmYear{2023}
\acmVolume{2023}
\acmNumber{2}
\acmDOI{XXXXXXX.XXXXXXX}
\acmISBN{}
\acmConference{Proceedings on Privacy Enhancing Technologies}
\begin{document}
\hyphenation{TikTok TikToks}
\title[Creative beyond TikToks: Investigating Adolescents' Social Privacy Management on TikTok]{Creative beyond TikToks: Investigating Adolescents' Social Privacy Management on TikTok}

\author{Nico Ebert}
\email{nico.ebert@zhaw.ch}
\orcid{0000-0002-9683-4792}
\affiliation{%
 \institution{Zurich University of Applied Sciences, School of Management and Law}
 \streetaddress{Theaterstrasse 17}
 \city{Winterthur}
 \state{Zurich}
 \country{Switzerland}
 \postcode{8401}
}

\author{Tim Geppert}
\affiliation{%
 \institution{Zurich University of Applied Sciences, School of Management and Law}
 \streetaddress{Theaterstrasse 17}
 \city{Winterthur}
 \state{Zurich}
 \country{Switzerland}
 \postcode{8401}
}

\author{Joanna Strycharz}
\affiliation{%
 \institution{University of Amsterdam, Faculty of Social and Behavioural Sciences}
 \streetaddress{Nieuwe Achtergracht 166}
 \city{Amsterdam}
 \state{North Holland}
 \country{Netherlands}
 \postcode{8401}
}

\author{Melanie Knieps}
\affiliation{%
 \institution{University of Zurich, Digital Society Initiative}
 \streetaddress{Rämistrasse 69}
 \city{Zurich}
 \state{Zurich}
 \country{Switzerland}
 \postcode{8401}
}

\author{Michael Hönig}
\affiliation{%
 \institution{Zurich University of Applied Sciences, School of Management and Law}
 \streetaddress{Theaterstrasse 17}
 \city{Winterthur}
 \state{Zurich}
 \country{Switzerland}
 \postcode{8401}
}

\author{Elke Brucker-Kley}
\affiliation{%
 \institution{Zurich University of Applied Sciences, School of Management and Law}
 \streetaddress{Theaterstrasse 17}
 \city{Winterthur}
 \state{Zurich}
 \country{Switzerland}
 \postcode{8401}
}

\renewcommand{\shortauthors}{Ebert et al.}

\begin{abstract}
TikTok has been criticized for its low privacy standards, but little is known about how its adolescent users protect their privacy. Based on interviews with 54 adolescents in Switzerland, this study provides a comprehensive understanding of young TikTok users' privacy management practices related to the creation of videos. The data were explored using the COM-B model, an established behavioral analysis framework adapted for sociotechnical privacy research. Our overall findings are in line with previous research on other social networks: adolescents are aware of privacy related to their online social connections (social privacy) and perform conscious privacy management. However, we also identified new patterns related to the central role of algorithmic recommendations potentially relevant for other social networks. Adolescents are aware that TikTok's special algorithm, combined with the app's high prevalence among their peers, could easily put them in the spotlight. Some adolescents also reduce TikTok, which was originally conceived as a social network, to its extensive audio-visual capabilities and share TikToks via more private channels (e.g., Snapchat) to manage audiences and avoid identification by peers. Young users also find other creative ways to protect their privacy such as identifying stalkers or maintaining multiple user accounts with different privacy settings to establish granular audience management. Based on our findings, we propose various concrete measures to develop interventions that protect the privacy of adolescents on TikTok.
\end{abstract}


\maketitle

\section{Introduction}
The global popularity and rapid growth of TikTok are accompanied by problems that are the subject of public debate. The platform has been criticized for several privacy issues such collecting personal data from minors under the age of 13 \cite{ftcVideoSocialNetworking2019, europeandataprotectionboardItalianDPAImposes} or transferring U.S. minor's data to China \cite{ryanTikTokPrivacyConcerns2020}. These issues are especially alarming as a large number of platform users are underage \cite{TikTokUsersAge}. As a result, the private space of children such as the bedrooms from which they create their videos becomes visible to the world \cite{kennedyIfRiseTikTok2020}. TikTok has reacted to public criticism by introducing several features to better protect the privacy of adolescents \cite{terripetersHowTikTokNew2021}. 

Implicit to the current debate is the apparent consensus that adolescents lack the awareness or skill set to consider the possible privacy implications of their platform use. In fact, however, little evidence is publicly available on how TikTok is used by adolescents \cite{montagPsychologyTikTokUse2021} and how they manage their privacy on the platform \cite{kangTeensPrivacyManagement2021}. As short videos are the platform's key purpose, this immediately raises the question of how younger users protect their privacy when they create videos. In this paper, we aim to answer the following research question: \textit{"How and why do adolescents manage their privacy when creating videos on TikTok?"} We build on the COM-B model for behavioral analysis, an established conceptual framework for behavior change widely applied in health communication and beyond \cite{michieBehaviourChangeWheel2011}. This model allows us to explore not only privacy behavior, but also related users' motivations, skills and desires. To explore adolescents' privacy management in video creation from the perspective of their capabilities, motivations, and opportunities, we conducted interviews with 54 adolescent TikTok users in Switzerland, where TikTok has gained popularity among young people \cite{bernathJAMESJugendAktivitaten2020}.

This study contributes to the growing body of research that addresses how adolescents manage their privacy on social networks. This topic, which is often viewed (and judged) through the moral lens of adults, is usually met with a sense of alarm. Empirical evidence that could serve as a better fact base about adolescents' online privacy behavior is largely based on platforms that cater to a more general population (e.g., Facebook, Twitter). However, TikTok is not only explicitly geared towards a younger audience \cite{leePopularMusicalLy2018}, but also strongly encourages the sharing of short personal videos. Although adding text is possible on TikTok, it takes a back seat in favor of video content. Given the particularly sensitive nature of one's image and its link to an individual's personal development \cite{europeancourtofhumanrightsFactsheetRightProtection2022}, TikTok strikes us as a particularly relevant but under-researched new use case with the potential to enrich the ongoing debate about how – if at all – teenagers perceive and manage their privacy \cite{kangTeensPrivacyManagement2021}. 

This study is the first to examine how adolescents between the ages of 12 and 18 manage their privacy on TikTok when it comes to personal videos. Our findings are based on original data from personal interviews and offer unique insights into how privacy concerns influence young people's online behavior. The qualitative nature of our study helped us to understand the components that shape sharing behavior on TikTok. Ultimately, this allowed us to make concrete suggestions on how to effectively promote privacy-protective behavior among adolescents on TikTok (e.g., specific training, improved app features, and policy enforcement). 

\section{Related Work}
\subsection{TikTok and Privacy Issues}
As with many social media platforms, TikTok has come under scrutiny for its handling of personal data. TikTok is a video-focused social network originally started as U.S.-based musical.ly but later bought by Beijing ByteDance Technology Ltd. The TikTok app (available for Android and iOS) allows users to create short videos (which may only be a few seconds long) and live streams \cite{kayeCoevolutionTwoChinese2021}. Like YouTube, TikTok is a manifestation of user-generated media where content is not primarily created by a limited number of producers but by a myriad of users \cite{khanSocialMediaEngagement2017}. Compared to other social networks such as Facebook or Instagram, users on TikTok do not need to communicate with each other to find a community. They can simply visit the “For You” default page to find like-minded users \cite{deleynInbetweenChildPlay2021, kayeCoevolutionTwoChinese2021,kennedyIfRiseTikTok2020, zulliExtendingInternetMeme2020}. Via the primary button at the center of the home screen, users can easily record and edit short videos, apply various effects and sounds, and reuse content produced by other users. Videos can be saved as drafts or published immediately to be viewed by different audiences (myself, followers, everybody) \cite{bytedanceltd.CreatingVideosTikTok}. As of September 2021, 1 billion monthly active users were reported \cite{wangTikTokHitsBillion2021}, and 740 million first-time installs were estimated in 2021 \cite{statistaTikTokAnnualInstalls}. Cloudflare, a provider of content delivery networks, ranked TikTok as the most popular website of 2021, before Google \cite{kalhanrosenblattTikTok}. TikTok is currently also gaining in popularity among users below the official age limit of 13 years \cite{demeulenaereOnderzoeksrapportDigitaleLeefwereld2020}. In Switzerland, three-quarters of all adolescents had a TikTok account in 2020 (behind Instagram and Snapchat with both over 90\%) \cite{bernathJAMESJugendAktivitaten2020}. Younger adolescents (12-15 years) were even more likely to have a TikTok account than older adolescents (16-19 years). Slightly more girls (78\%) used it than boys (68\%). 51\% of all adolescents stated to use it at least multiple times per week, and 38\% daily \cite{bernathJAMESJugendAktivitaten2020}. However, little is known about how young users think about the data they share on the platform.

The app has raised numerous severe security and privacy concerns (e.g., \cite{wongTikTokAccusedSecretly2019, eberlPrivacyAnalysisTiktok2019, bergmanMajorTikTokSecurity2020,doddsHowPopularApps2020, krauseIOSPrivacyAnnouncing2022}) and caught the attention of the international authorities in the U.S. and EU \cite{europeandataprotectionboardItalianDPAImposes, ExclusiveOpensNational2019, ryanTikTokPrivacyConcerns2020}. For example, an analysis of the app revealed extensive aggressive user tracking (e.g., including techniques such as fingerprinting) and data sharing with other websites (e.g., sharing searches with Facebook) \cite{eberlPrivacyAnalysisTiktok2019}. The app could also potentially collect other personal data from the user's smartphone (e.g., data from the clipboard \cite{doddsHowPopularApps2020}). Since young people have always been an important user group of TikTok, concerns have been raised about ByteDance's handling of their personal data. For example, in February 2019 ByteDance was fined USD 5.7 million by the U.S. Federal Trade Commission (FTC) because musical.ly had collected information from minors under the age of 13 in violation of the Children's Online Privacy Protection Act \cite{ftcVideoSocialNetworking2019}. Due to the death of a 10-year-old TikTok user, the Italian data protection authority has banned TikTok from processing the data of users whose age could not be determined with full certainty \cite{europeandataprotectionboardItalianDPAImposes}. Also, the transfer of minors' data to China after the acquisition of U.S.-based musical.ly had caused a serious backlash in the US and EU \cite{ExclusiveOpensNational2019, ryanTikTokPrivacyConcerns2020}. As recent as June 2022, evidence surfaced that ByteDance has repeatedly accessed U.S. user data from China – a practice that they had denied three years earlier when similar criticism was raised \cite{emilybaker-whiteLeakedAudio80}. 

TikTok has reacted to public criticism with several privacy-related updates to the original app. As part of its settlement with the FTC, the platform introduced an age-verification process for its users based on self-declaration, meaning users can provide a false age \cite{leeTikTokStopsYoung2019}. Further changes included extended parental control features \cite{kastrenakesTikTokNowLets2020} and privacy settings contingent on the app users' age statement \cite{terripetersHowTikTokNew2021}. While children below 13 cannot use the app, adolescents between the ages of 13 and 15 are automatically switched to a “private account” as a default option, limiting those who can view their videos to approved followers. When 16- and 17-year-old users imitate an existing video in the form of a “duet” (split-screen video) or “stitch” (video incorporating a short clip of someone else's content), these are automatically restricted to “friends only”. Only users who are 18 and older can buy and send virtual gifts. However, it is unclear if and how TikTok's efforts have affected users' privacy management.

\subsection{Adolescents' Privacy Management on Social Media}
From the moment adolescents started to use online social networking sites, “online privacy” has been a major topic of discussion \cite{fengTeensConcernPrivacy2014}. Informational privacy can be defined as “the claim of individuals, groups or institutions to determine for themselves when, how, and to what extent information about them is communicated to others” \cite{westin1967privacy}. 
Research on online privacy and adolescents can be divided into two categories: “institutional privacy” and "social privacy" \cite{raynes-goldieAliasesCreepingWall2010}. Institutional privacy refers to the data collection practices by organizations (e.g., for commercial purposes) \cite{raynes-goldieAliasesCreepingWall2010, youngPrivacyProtectionStrategies2013}. The focus of this paper is social privacy, i.e., issues related to sharing personal information with others (e.g., friends and family).
According to the theory of "networked privacy," individuals do not have complete control over the sharing of their personal information within social connections (e.g., on social media) because privacy is not managed by individuals alone, but by networks of individuals collectively \cite{marwickNetworkedPrivacyHow2014}.

 Young people are often seen as particularly vulnerable social media users with limited capacities to protect their privacy \cite{deleynReframingCurrentDebates2019, marwickNetworkedPrivacyHow2014}. At the same time, they are also portrayed as individuals who put themselves and others at risk with their naive and reckless social media behavior \cite{dewolfControlResponsibilityDiscursive2019}. Following this logic, numerous guides for parents emphasize the importance of modifying privacy settings and monitoring their children's behavior (e.g., \cite{hodgeParentGuideTikTok2020}). However, there has also been a pushback to this alarmist perspective by scholars who suggest that adolescents' online privacy should be addressed based on empirical research rather than paternal instinct \cite{wisniewskiPrivacyAdolescence2022}.

Empirical evidence from social networks other than TikTok (e.g., Facebook) suggests that adolescents are aware of their social privacy and actively manage their privacy on social media. As described by boyd \cite{boydItComplicatedSocial2014}, adolescents want to avoid surveillance from parents, teachers, friends and other meaningful persons in their lives (that is what “online privacy” means to them). Adolescents' social media use seems to generally prompt increased disclosure of personal information \cite{shinAdolescentsPrivacyConcerns2016}. However, frequent sharing of content does not imply that adolescents share indiscriminately, nor that the content is intended for a wider audience \cite{marwickNetworkedPrivacyHow2014}. Indeed, adolescents are concerned about their privacy and capable of protecting it \cite{balleysBeingPubliclyIntimate2017, blank2014new, dewolfGroupPrivacyManagement2016, livingstoneChildrenDataPrivacy2019}. Contrary to conventional wisdom, young people are, in fact, more likely to protect their privacy on social media than older people \cite{blank2014new}. Madden et. al found several strategies adolescents use on social media to manage their identity and protect sensitive information \cite{maddenTeensSocialMedia2013}. These strategies include deleting friends, faking names, deleting content, withholding/faking information, and changing privacy settings \cite{dennenContextCollapseStudent2017, heirmanOpenBookFacebook2016, mullenAdolescentsResponseParental2016}. They also employed different “zones of privacy” by using different channels for disclosing personal information to maintain intimacy with friends while protecting their privacy from their parents and strangers \cite{livingstone2008taking}. Privacy management can also mean modifying social media content to shield it from audiences \cite{maddenTeensSocialMedia2013, mullenAdolescentsResponseParental2016}. This practice is referred to as “social steganography” or encoding a message for a defined audience \cite{marwickNetworkedPrivacyHow2014}. Adolescents' privacy management is influenced by various factors such as their social environment (e.g., friends, parents), prior (negative) experiences as well as the saliency of privacy settings \cite{livingstoneChildrenDataPrivacy2019, zaroualiEverythingControlPrivacy2018}. 

Despite the existing evidence on adolescents' social media use on other social networks, researchers argue that existing findings might not be directly applicable to TikTok \cite{montagPsychologyTikTokUse2021}. Compared to other networks such as Facebook or Instagram, TikTok mainly thrives on content exploration and (re)-creation \cite{zulliExtendingInternetMeme2020}. The focus is not on the interaction between users and their social network but the interaction with users' videos proposed by an algorithm \cite{bhandariTiktokAlgorithmizedSelf2020}. The main feature, the “For You” page, presents an endless stream of personalized, publicly available videos. Seeing them will motivate users to react and create similar content (e.g., through features such as “duet” or “stitch”). TikTok might therefore pose a particular threat to adolescents' privacy because a space previously conceptualized as private and safe can easily become a space of public visibility, surveillance, and judgment (such as in the case of a teenager being seen to perform a dance routine in their bedroom) \cite{kennedyIfRiseTikTok2020}.

Only a few studies have investigated adolescents' privacy management on TikTok. There is some evidence that privacy management on TikTok is considered as crucial by adolescents \cite{deleynInbetweenChildPlay2021} and becomes more stringent at higher perceived risks \cite{kangTeensPrivacyManagement2021}. However, it is unclear how and why adolescents manage their privacy on TikTok. 

\subsection{COM-B Model}
As we were interested in the components that shape privacy behavior, we chose the COM-B model, which has been used in exploratory studies (e.g., \cite{flannery2018enablers}) and a series of contexts to change behavior (e.g., \cite{barkerApplyingCOMBBehaviour2016}), as the conceptual framework for our analysis. Many behavioral theories have been developed, often with overlapping but differently named \mbox{constructs \cite{michieBehaviourChangeWheel2014}} and limited guidance on choosing an appropriate theory for a particular, real-world context \cite{michieBehaviourChangeWheel2011}. As a consequence, theories are often under-used to understand real-world contexts and to design real-world solutions, which makes replication, implementation, evaluation, and improvements difficult \cite{ecclesExplainingClinicalBehaviors2012, michieBehaviourChangeWheel2011}. Researchers have argued that a comprehensive meta‐model or “supra‐theory” model of behavior – like the COM-B model – is needed that is applicable across contexts \cite{ecclesExplainingClinicalBehaviors2012, michieBehaviourChangeWheel2011}. As a meta-model of behavior, the COM-B model does not come with a pre-determined set of context-specific predictions that are common for many behavioral theories. COM-B is based on several existing social cognition models and has a broader understanding of behavior, having "also [...] automatic processing at its heart [like emotions and habits], broadening the understanding of behaviour beyond the more reflective, systematic cognitive processes that have been the focus of much behavioural research [...] (for example, social cognition models such as the Theory of Planned Behaviour)" \cite{michieBehaviourChangeWheel2011}. Its comprehensive nature and flexibility made it a good fit for the exploratory nature of our study that was not constrained by the conceptual boundaries of a single theoretical framework. Furthermore, the model comes with hands-on actionable advice on appropriate interventions in a given context in form of a holistic behavior change framework (“Behavior Change Wheel”) (see \cite{michieBehaviourChangeWheel2011}).

\begin{figure}[t]
 \centering
 \includegraphics[trim=43 0 0 0,clip,width=1\columnwidth]{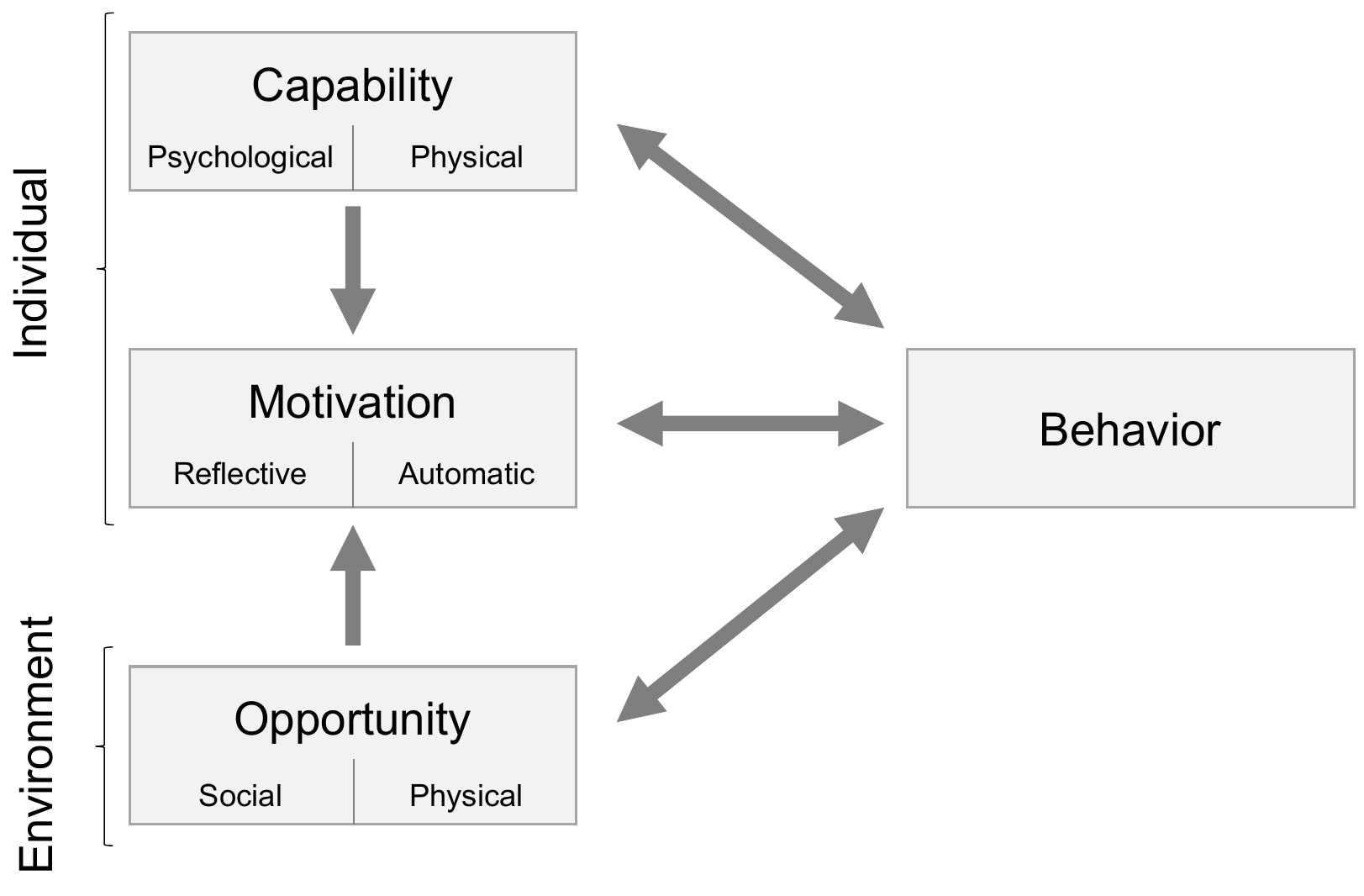}
 \Description{The figure depicts the COM-B model with its four dimensions capability, motivation, opportunity and behavior}
 \caption{The COM-B model \cite{michieBehaviourChangeWheel2011}. \textnormal{ The three components capability (C), opportunity (O) and motivation (M) must be present for a behavior (B) to occur. They interact over time and form a dynamic system with positive and negative feedback loops \cite{west2020brief}.}}
 \label{fig:com-b}
\end{figure}

As illustrated in Figure \ref{fig:com-b}, the COM-B model is based on three components – capability (C), opportunity (O), and motivation (M) – that shape a person's behavior (B) \cite{michieBehaviourChangeWheel2011}. Firstly, capability is a subject's psychological ability (including necessary comprehension, knowledge, and skills) as well as the physical ability (e.g., control of the body) to engage in a behavior. Secondly, motivation can be defined as the subject's mental processes that energize and direct behavior. It includes the reflective motivation that involves conscious processes (e.g., goals, plans, and evaluations) as well as automatic processes (i.e., habitual, instinctive, drive-related, and affective processes). Finally, opportunity is defined as an attribute of the environmental system (unlike capability and motivation) that enables or facilitates a behavior. Opportunities can be physical (e.g., technical features of an app, material, financial, and time) and social (e.g., norms and culture). In this study, we analyzed the participants' capabilities, opportunities, and motivation to engage in privacy behaviors.

Firstly, capability is a subject's psychological ability (including necessary comprehension, knowledge, and skills) as well as the physical ability (e.g., control of the body) to engage in a behavior. Secondly, motivation can be defined as the subject's mental processes that energize and direct behavior. It includes the reflective motivation that involves conscious processes (e.g., goals, plans, and evaluations) as well as automatic processes (i.e., habitual, instinctive, drive-related, and affective processes). Finally, opportunity is defined as an attribute of the environmental system (unlike capability and motivation) that enables or facilitates a behavior. Opportunities can be physical (e.g., technical features of an app, material, financial, and time) and social (e.g., norms and culture). 

In our exploratory study, we did not focus on identifying interactions between COM-B components that explain a specific target behavior. Rather, our scope was first to learn about the full range of behaviors and explanatory factors associated with adolescents’ privacy management. 

\section{Methodology}
\subsection{Research Ethics}
This paper is based on semi-structured, one-to-one interviews with adolescents in the Canton of Zurich, Switzerland, conducted in November 2021. In total, we visited two secondary schools (one in the city of Zurich and one in the greater Zurich area) and three youth centers (all in the city of Zurich). All interviews were audio-recorded and transcribed verbatim. Ethical approval was obtained from our university's institutional review board. Study participants provided written informed consent. For subjects below the age of 16, additional consent was sought from the parents. The interviews were voluntary and conducted at the institutions from which the subjects had been recruited. Digital, personalized shopping vouchers with a value of CHF 20 (\textasciitilde USD 19) were offered to study participants as compensation. The amount and type of the vouchers was determined beforehand together with the adolescents' supervisors (i.e., teachers, social workers) in order to not create an inappropriate but still sufficient incentive. After the interviews, in agreement with the participants, WhatsApp was used to deliver the individualized vouchers to the participants and to allow them to review their personal interview transcripts.

Several steps were taken to protect the participants identity without compromising the transparency of our research process. To begin with, all personally identifiable information was removed (e.g., references to persons, locations) and participants' names were replaced with pseudonyms. Furthermore, the study data was stored in line with our university's storage policy and only the involved researchers had access to the files. Finally, the original audio files were deleted from all devices half a year after recording together with other remaining personal data (e.g., phone numbers, WhatsApp chats, digital vouchers). 

\subsection{Sample and Procedure}
Due to the lack of research on this topic, we chose a highly exploratory approach. To identify information-rich cases and make optimal use of available resources, we drew a purposive sample \cite{etikan2016comparison}. We used social media and search engines to find institutions in the Canton of Zurich (e.g., secondary schools, youth work, youth associations, museums) with contact with adolescents between 12 and 18 years of age. Afterward, principals of participating institutions recruited interested teachers and social workers. They, in turn, contacted interested TikTok users in the required age group. To extend the participant base, we applied snowballing among the interested TikTok users. Based on our primary aim (i.e., to explore how adolescent TikTok users manage their privacy), we chose to sample based on study participants' age and gender (equally distributed). We decided to ignore other demographic information such as ethnic identity. Following a pragmatic definition of theoretical saturation \cite{lowPragmaticDefinitionConcept2019}, no new information emerged after approximately 40 interviews, and we ended data collection after the 54th interview.

We chose to employ semi-structured interviews for our study because it encourages two-way communication and provides the interviewer with the opportunity to learn the reasons behind an answer. Some of the questions were part of the interviewer's guide (see Appendix), others were addressed at the moment. The interview guide was developed based on a previous study that applied the COM-B model in a qualitative setting \cite{castroUsingBehaviorChange2021} and adopted to the context of the current study. After asking for demographic information, we first explored general TikTok usage and motivation. The other questions followed the COM-B structure and were related to privacy-related behaviors as well as the explanatory components related to the target behavior “video creation”. We finished the interviews with questions about commercial privacy aspects (i.e., targeted advertising and user tracking). After the interview process was completed, study participants received a copy of their interview transcript via \mbox{WhatsApp} and were invited to add information or make amendments. Minimal revisions were made by one participant. 

To analyze the content of the interviews, we used a two-step procedure that first divided each statement into one of the four COM-B components (behavior, capability, opportunity, motivation) before further subdividing them into privacy-specific content. For phase one, we used a directed content analysis approach \cite{hsiehThreeApproachesQualitative2005} to analyze the statements. To counter the subjectivity inherent to qualitative data analysis, three researchers read and coded all statements into the four COM-B domains (behavior, capability, opportunity, motivation). On the grounds of economy in both cost and effort, we decided against using "intercoder reliability" (ICR). As full replication of results was deemed unnecessary due to the exploratory and qualitative nature of data collection and analysis, we instead followed guidelines suggesting the use of "multiple coding" which allows independent researchers to cross check their coding strategies and interpretation of data \cite{barbour2001checklists}. The authors engaged in researcher triangulation \cite{denzin2017sociological} by discussing the emerging codes during the open coding process of the first three interviews and developed coding guidelines. Disagreements were discussed and resolved. Using the MAXQDA 2020 software, all responses were coded consistent with six COM-B labels\footnote{We did not need to code “physical capability” as the participants did not have physical impairments.} (behavior, psychological capability, automatic/reflective motivation, social/physical opportunity). To ensure continued adherence to the agreed coding guidelines, the three researchers regularly communicated to ensure coding consistency.

In phase two, all statements – previously labeled as one of the COM-B components – were further analyzed for their privacy-specific content. Therefore, an inductive thematic analysis \cite{braunUsingThematicAnalysis2006} to identify themes within similarly coded statements was conducted (see Appendix for coding scheme). One researcher identified themes across identically coded statements and discussed them with the other researchers. A theme reflects a collection of similar responses from at least two different study participants. For example, responses that were coded under the COM-B label “reflective motivation” such as “I would be afraid of stupid remarks.”, “I have no desire to be bullied.”, and “I can do without being ridiculed in my class's WhatsApp group.” were allocated to the privacy-specific theme “negative reaction avoidance”. This step resulted in a list of themes within each of the six COM-B labels. Ultimately, the researchers reviewed and discussed the emerging themes, merged similar themes, and re-labeled others. By playing the “devil’s advocate” – a common way to scrutinize identified themes \cite{barbour2001checklists} – we sought to exploit the full potential of multiple coding to furnish alternative interpretations of our findings. The anonymized, coded interview transcripts are publicly available at \href{https://osf.io/z8d3w/}{osf.io/z8d3w}. 

\section{Results}
A total of 54 adolescents aged between 12-18 years (15 ± 1.82 years) were interviewed, of which half (27) were female (see Table \ref{tab:descriptivestatistics}). Interviews ranged from 5 to 21 minutes in length, with a mean of 12.6 min per interview (SD = 3.91). Most users attended secondary school, and 80\% had used the app for more than one year. Half of the study participants admitted using TikTok between one and three hours per day. 
\begin{table}[h]
 \caption{Characteristics of one-to-one interview participants (\textit{n} = 54)}
 \label{tab:descriptivestatistics}
 \begin{tabular}{lc}
 \hline
 Variables&\% (\textit{n})\\
 \hline
 Gender (\% of females)&50\% (27)\\
Age&15 ± 1.82\\ 
Educational level& \\	
\hspace{2mm}Primary level&2\% (1)\\
\hspace{2mm}Lower secondary level&54\% (29)\\
\hspace{2mm}Upper secondary level&44\% (24)\\
User since& \\
\hspace{2mm}One year or less&20\% (11)\\
\hspace{2mm}Between one and two years&33\% (18)\\
\hspace{2mm}More than two years&46\% (25)\\
Current app usage& \\
\hspace{2mm}Daily >= 3h&17\% (9) \\
\hspace{2mm}Daily >= 1 and <3h&50\% (27) \\
\hspace{2mm}Daily < 1h&28\% (15) \\
\hspace{2mm}Less than daily&6\% (3) \\
 \hline
\end{tabular} 
\end{table}

Building on the conceptional framework of the COM-B model, we identified 13 themes from the data analysis that described how and why adolescents protect their privacy on TikTok (see Table \ref{tab:conceptualframework}). These are described in more detail in the following. No weighting was associated with the themes in terms of their overall contribution. 
\begin{table*}[t]
 \caption{Identified themes for adolescents' video privacy management on TikTok based on the COM-B model. Frequency is calculated across 54 interviews.}
 \label{tab:conceptualframework}
 \begin{tabular}{lp{10cm}c}
 \hline
 Theme&Description&Frequency\\
 \hline
\textit{Behavior}&&\\	
\hspace{2mm}Proactive privacy&Publishing videos with control over the content and the audience&30\\
\hspace{2mm}Avoidance&Publishing no videos on the platform&24\\ [0.5ex]
\textit{Capability (Psychological)}&&\\	
\hspace{2mm}Past privacy incidents&Previous negative experiences related to privacy on the platform (e.g., lost account, accidental publication)&15\\
\hspace{2mm}Privacy literacy&Knowledge and skills related to privacy management in the app (e.g., audience understanding and configuration)&53\\ [0.5ex]
\textit{Opportunity (Social)}&&\\ 	
\hspace{2mm}Negative feedback&Negative behavior of others affects privacy management (e.g., observation of cyber-bullying)&16\\ 
\hspace{2mm}Linkability experience&Observing that online personas can be linked to the personal sphere affects privacy management (e.g., my teacher is on the platform)&39\\ 
\hspace{2mm}Restrictive influence&Restrictive behavior of others affects privacy management (e.g., restrictive parental mediation)&34\\ [0.5ex]
\textit{Opportunity (Physical)}&&\\	
\hspace{2mm}Platform features&Privacy-related features of the platform (e.g., audience settings, sharing via other social networks)&46\\ 
\hspace{2mm}Device features&Privacy-related features of the device (e.g., screen time limits, deleting videos on the smartphone)&17\\ [0.5ex]
\textit{Motivation (Automatic)}&&\\	
\hspace{2mm}Negative emotion avoidance&Avoidance of negative emotions expected as a result of publication (e.g., shame, fear)&15\\ [0.5ex]
\textit{Motivation (Reflective)}&&\\	
\hspace{2mm}Negative reaction avoidance&Goal to avoid expected negative consequences of publication&10\\
\hspace{2mm}Privacy identity&Privacy as a general value (e.g., also on other platforms)&5\\ 
\hspace{2mm}Publicity avoidance&Goal to avoid expected publicity of publication&29\\ 

 \hline
\end{tabular} 
\end{table*}

\subsection{Behavior}
\subsubsection{Proactive privacy}
The participants in our study mentioned various ways to control the content of their TikToks\footnote{The term “TikToks” is used synonymously with videos.} and their audience. Publishing content to audiences was described as reflective and non-automatic (as opposed to a habitual, non-reflective publication of TikToks). This behavior is also referred to as the “approach” privacy strategy \cite{marwickNetworkedPrivacyHow2014}. For example, regarding the content, study participants described what they consider to be too sensitive for publication on TikTok and would not publish (e.g., TikToks that reveal too much about them). Lima (F, 14) creates public videos and has 50 different accounts. She has clear privacy boundaries regarding the video content: “I would not post TikToks where you can see a lot of myself. I wouldn't post videos in which I'm drunk.”. Another form of restriction is to define who can see which type of content on the platform. This includes TikTok users making drafts only visible to themselves or blocking selected users from watching videos. Bärbel (F, 13) actively tries to keep her parents from seeing her videos: “To prevent my parents from seeing my videos, I can simply block them.”.

We identified two subthemes within the proactive privacy theme: private creators (19 persons, 35\% of the sample) and public creators (11 persons, 20\%). Private creators create videos only for themselves or close friends but do not publish them for a broad audience. A few users described the practice of posting videos that are just visible to themselves, only to be able to then repost them on “more private” social media such as Snapchat or WhatsApp for a selected group of people: “I don't post my videos. I download them, save them under photos, then send them on WhatsApp, for example. I only use TikTok for editing.” (Amy, F, 17). 

Public creators regularly create videos for their followers or the general public. An extreme case is Joy (F, 13), who has used TikTok since she was nine years old (when the app was still musical.ly). She maintains 50 thematic user accounts with different age settings and distinct followings (e.g., some accounts for gaming-related videos and others for YouTube reposts). In addition to managing multiple accounts, public creator Lima (F, 14) also uses the live feature. It is available to users with at least 1,000 followers and allows them to create personal live streams and interact with users in real time. Lima had to set her age to 16 years to enable the live feature.
\subsubsection{Avoidance}
Some study participants reported that they do not publish videos on TikTok at all to protect their privacy. In the literature, this is referred to as the avoidance privacy strategy \cite{marwickNetworkedPrivacyHow2014}. Peter (M, 14), one of 24 study participants (44\%) we classified as a pure consumer, stated: “I've never created a TikTok. I don't even know how to do it.”. Tim (M, 12) published once but decided to only watch TikToks afterward: “To try it out, I uploaded something once, but nothing from me. I thought that was funny. But I prefer to watch videos.”
\subsection{Capability (Psychological)}
\subsubsection{Past privacy incidents}
This theme refers to a specific form of privacy-related knowledge (cp. \cite{michieBehaviourChangeWheel2014}) gained after experiencing potential or actual privacy incidents. Potential privacy incidents are perceived as minor threats but may lead to increased privacy awareness. “I posted my very first video by accident. It was only seen by three people,” reported Yasmina (F, 15). Lima (F, 14), a public creator, remembered: “I was half asleep and accidentally posted a TikTok. The next morning, I saw that someone had commented on the video. But I thought it was funny and not bad at all.” When TikTok updated its app and increased the size of the “publish” button to lower the threshold for publication, Lima decided to block app updates. 

Users have also realized that some of TikTok's privacy features can be easily bypassed. Their awareness of the platform's weaknesses has contributed to a greater privacy awareness. An example is a feature that allows blocking certain users from viewing videos, which can be easily bypassed: “If I block people but they still want to see my TikToks, they immediately make an extra fake account and continue seeing them.” Roswitha (F, 15). However, she found a way to manage her privacy: “Since these users have too few followers, I simply block them again or ignore them depending on the video.”.

A more serious subtheme are actual privacy incidents. Bärbel (F, 13) had to realize that she was not anonymizing herself sufficiently: “I wore a mask on my face in the video, anonymously, so to speak. But the people who deal with me every day recognized me by my outfit, my room, and my hairstyle and posted the video in the class WhatsApp chat.”. Anna (F, 14) reported losing her account and not being able to reclaim it through TikTok's customer support. At the same time, other users were still able to watch her videos: “I made videos of myself when I was 9 and then lost the account. Now the videos are still public, but I can no longer access them.”. 
\subsubsection{Privacy literacy}
Privacy literacy can be defined as a combination of factual or declarative ('knowing that') and procedural ('knowing how') knowledge about online privacy \cite{treptePeopleKnowPrivacy2015}. Concerning the publication of videos on the platform, adolescents need to have the knowledge and skills to assess and manage audiences and content as needed. 

Respondents mentioned, for example, that the algorithm might present a video on TikTok's center stage: “It depends on how popular a video is and only then does it appear on the For You Page.” (Bärbel (F, 13)) or that public videos can also be watched without having a TikTok account: “From Google or Safari you can type in TikTok and view the videos.” (Aron (M, 13)). They also described how to find out which of their peers used TikTok: “When you post a video, it spreads immediately and then you know who has TikTok and who does not. Because so many people have TikTok now, it has become weird for me to post TikToks.” (Elsa (F, 14)). 
Respondents also described their audience and content management skills. The private creator Bea (F,14) only publishes for a strictly curated list of followers and therefore has established an approval process that allows her to maintain the desired level of privacy: “I get to know new classmates first and only then give them my TikTok account. Afterward, they tell me they sent a request and I accept them as followers in the app.” (Bea (F, 14)). Furthermore, the adolescents interviewed were also able to assess different levels of sensitivity of content in terms of their privacy and select an adequate audience accordingly: “My buddy and I made 10 TikToks in which we share our weekend activities with people. Some have 60,000 views. But we think carefully what to make public.” (Alex (M, 18)).

The adolescents also talked about various app settings needed to manage the audience, such as the activation of the private account “Switching to the private account takes only two minutes. This is not difficult.” (Alexandra (F, 12)) or knowing the publication status of a video: “A draft is rendered greyish and blurry. When published, it is bright and jumps right out at you.” (Alexander (M, 15)). Some adolescents also perform “digital housekeeping” activities by removing content related to a specific event or as a habit: “As I became older, I started to delete old videos.” (Ariane (F, 15)).
\subsection{Opportunity (Social)}
\subsubsection{Negative feedback}
Negative feedback refers to expected or observed negative feedback from others (such as harsh comments to videos). Study participants reported negative reactions on the platform (e.g., from strangers or people from the same school) as an explanation for their privacy protection behavior. Alexander (M, 15) mentioned a general culture of mutual criticism: “Many of the famous TikTokers sometimes make mistakes. Afterwards, everyone makes fun of them in videos.”. Other respondents mentioned negative reactions from their peers that had influenced their behavior: “A friend went viral with a video. Then she got yelled at on the street. It would annoy me.” (Katja (F, 17)).
\subsubsection{Linkability experience}
Similar to the perception of negative feedback, the realization of how easily online personas can be linked to the personal sphere can also lead to more restrictive publication behavior. Study participants perceived the platform as a public space shared by acquaintances and strangers. However, by recognizing people from their school on their “For You” page, study participants realized that they, too, could be easily recognized. As Georg (M, 15) put it: “There are maybe ten or twenty people in the school building who do [public] TikToks regularly. You suddenly realize: I know that guy from TikTok. That's the reason why I don't publish.”. In addition to peers, respondents also described experiences that made them understand that acquainted adults in authority positions would be able to see their TikTok as well. Sibylle (F, 15) realized this: “My music teacher was on TikTok singing a song.”. Therefore, Sibylle also does not publish so as not to be recognized by everyone on the platform.
\subsubsection{Restrictive influence}
Restrictive influence refers to others (e.g., close friends or parents) perceived to be restrictive or restricting study participants' video creation behavior. Some interviewees reported that their friends did not publish on TikTok, which in part motivated why they did not publish, either. In mentioning his peers, Felix (M, 12) stated: “Most of the people I know don't upload anything of themselves where they show their face.”. Another example is restrictive mediation by parents or relatives: “My eight-year-old cousin accidentally posted a video with my smartphone. His uncle saw it on his For You page, so I deleted it.” (Sibylle (F, 15)).
\subsection{Opportunity (Physical)}
\subsubsection{Platform features}
Age verification is a key platform feature intended to protect the privacy of young users (not limited to creating videos) and the subject of much public discussion. In the semi-structured interviews, 29 of the interviewed participants were also asked what age they provided. Two-thirds admitted that they had given a false age when they registered (indicating, e.g., the age of their parents). The main motivation for this behavior was to be able to use TikTok in general (for those below the age of 13) or all its features. Some study participants, like Martin (M, 14), also had misconceptions about possible age restrictions: “Because otherwise, TikTok won't let me watch videos.”.

However, study participants also described how they used TikTok's features for privacy purposes in general. This includes using a nickname instead of their real name, limiting the use of personal information on their profile page, and not linking their TikTok account with other social media accounts (e.g., Instagram). While some interviewees do not use a name at all: “Why should people know my name? I have replaced my name and individual letters with an X.” (Ali (M, 12)), others actively involve their parents to make use of the in-app parental controls that restrict their app access. 

Study participants also reported using various features related to audience configuration, such as creating personal drafts, activating a private account, deleting videos, or blocking users. Public creators sometimes create multiple “privacy-tailored” user accounts with specific follower groups for content of special sensitivity. Where the features offered by the platform are perceived as too limited or ineffective, the adolescents used creative workarounds not originally anticipated by the platform provider. For example, it is not easily possible to download and share drafts of videos that are not yet published. Amy (F, 17), however, described a popular workaround: “I post videos on TikTok, but only for me. Afterward, I'm able to download them to share them with my friends on WhatsApp.”
\subsubsection{Device features}
As part of the greater sociotechnical system, some devices (e.g., smartphones) offer features that affect user privacy. For example, study participants make use of the “digital wellbeing” functionality of their smartphone to limit their screentime: “I used TikTok three hours a day because I didn't know anything better to do with myself. Now I'm trying to get a handle on this with a screen time limit.” stated Matthias (M, 17). Sandra (F, 14) was one of the study participants who used smartphone features to share videos more selectively: "You can take a screenshot of drafts with an iPhone and then send them via WhatsApp or Snapchat.". As mentioned earlier, Lima (F, 14) noticed that the size of the red “publish” button grew with each new app update compared to the grey “save as draft” button. Fearing accidental publication, she bypassed this potentially manipulative design pattern (“dark pattern”) by using an old version of the app, which her operating system allowed her to do: “Therefore, I have blocked the updates for TikTok on my cell phone.”.
\subsection{Motivation (Automatic)}
\subsubsection{Negative emotion avoidance}
The interviewees describe various negative emotions if they appeared in a video on TikTok. For example, they mentioned feelings of discomfort, shame, awkwardness, and annoyance. Milo (M, 12), who does not publish any videos, said: “I would be embarrassed to be seen in a video.” Elsa (F, 14) reported that her desire to avoid negative emotions had evolved. While she had posted videos on musical.ly, she didn't publish on TikTok anymore: “Posting TikToks has become weird for me.”.
\subsection{Motivation (Reflective)}
\subsubsection{Negative reaction avoidance}
Another reason for not publishing personal content was negative reactions by others to their videos such as being bullied in class (e.g., in the WhatsApp class chat). Alexander (M, 15), who does not publish any videos, commented: “You make a mistake, people from school see it, it gets sent on, and you get bullied.”. Avoidance can also relate to the negative long-term consequences of sharing personal content. As they get older, adolescents who are getting ready to join the job market realize that their activity on TikTok could harm their career prospects. “The Internet never forgets and if I eventually look for an apprenticeship, it may be that my future employer sees that. That's very bad for my reputation.” (Lima (F, 14)).
\subsubsection{Privacy identity}
With privacy identity, we refer to a coherent set of privacy-related behaviors and personal qualities of an individual in a social setting \cite{michieBehaviourChangeWheel2014}. Some teenagers consider privacy as a value in itself and part of their identity. For example, for Yara (F, 14), the publication of videos on TikTok is no different from any social network activity: “It's just not my thing. I don't post in general either, not even on Instagram or anything.”. Lena (F, 17) explicitly stated that she considers privacy a significant personal value: “Privacy is important to me. I keep everything private that can be kept private.”. 
\subsubsection{Publicity avoidance}
Another motivation for restricting the publication of personal videos on the platform is closely related to the linkability experience theme: the desire to not attract public attention. Study participants explained that publishing on TikTok means being in the public eye: “It's a big platform, and I don't want people around me to see that I make videos.” (Anna (F,14)). While in musical.ly, the public was described as a community of people with similar interests and ages, on TikTok, it is perceived as a heterogenous, superficial place with different people of all ages (including strangers, peers from the same school, teachers, extended family members, and parents). Lina (F, 17) described how the change in the audience had an impact on her behavior: “At musical.ly, there were also strangers, but more my age. But TikTok is now worldwide and there are adults everywhere. I don't have to post anything there.”. Her comment shows that the platform is now perceived as completely public, whereas it used to be a more private community.

\section{Discussion}
Our general observation of adolescents' on TikTok is in line with previous research on other social networks \cite{balleysBeingPubliclyIntimate2017,blank2014new,dewolfGroupPrivacyManagement2016,livingstoneChildrenDataPrivacy2019}: Contrary to public perception which portrays the publication of TikToks by young people as automatic and unreflective, the adolescents in our sample actively engaged in privacy management. They demonstrated a strong awareness of the need to manage their online identity and social privacy on the platform. However, the interview participants were more concerned with protecting their privacy from their immediate social environment than with institutional or commercial privacy issues. That is, while they were generally aware that TikTok used algorithms to tailor video content to their particular online behavior, they were more worried about the tangible aspects of the algorithm: that a published video could immediately appear on a classmate's account. 

Next, we will discuss the results in more detail following the structure of the COM-B model. While many of our findings are consistent with themes found in previous research on other social media platforms (e.g., Facebook), a few themes and aspects are indeed unique and – best to our knowledge – have not yet been studied by researchers on TikTok or other platforms. The qualitative nature of our data inform the design of very concrete interventions on TikTok (Section \ref{section:intervention}).  

\subsection{Behavior}
In addition to previous research on other social networks \cite{marwickYouthPrivacyReputation2010}, we were able to identify two very different types of proactive privacy behavior: public and private creation. While public creators perform privacy management to share videos directly on TikTok, private creators merely use the platform to create and edit videos to share them on other social networks that they see more appropriate for such content (e.g., Snapchat, WhatsApp). It indicates that adolescents have different "imagined audiences" (mental conceptualization of the people with whom the user is communicating, \cite{eden2012knock}) on each social network and curate who sees what by switching between networks. A unique finding of our study is that private creators essentially reduce TikTok, which was originally conceived as a social network, to its extensive audio-visual capabilities and share their personal content where social connections already exist and a higher degree of perceived control and intimacy exists (e.g., WhatsApp). It is possible that such a practice might also be found elsewhere (e.g., Instagram, YouTube). At a time when adolescents' increasingly use multiple social media platforms at once, privacy perceptions of and management between different platforms has to be addressed more comprehensively. That is, privacy management can no longer be seen as a single-platform-phenomenon – an observation with important research implications. Rather than focusing on isolated social networks with their own privacy standards, researchers should expand their analysis to include a cross-network view of privacy management.

\subsection{Psychological Capabilities}
Similar to previous studies on other social media platforms \cite{livingstoneBalancingOpportunitiesRisks2010, balleysBeingPubliclyIntimate2017, livingstoneChildrenDataPrivacy2019}, we found that adolescents possess knowledge and skills on how to manage their privacy on TikTok (see "privacy literacy" theme). That is, adolescents were not only able to assess the audience of videos but also to actively manage the audience and content of their TikToks. 
As previously noted \cite{marwickNetworkedPrivacyHow2014}, privacy management can be very creative. This finding also holds true for TikTok: some of our respondents reported using various accounts for different audiences, blocking app updates to avoid receiving less privacy-friendly versions of the app, and making an effort to detect fake users trying to follow them. An interesting observation that can potentially inform other research on social privacy management in social networks is that adolescents on TikTok do not only use the technical features provided by the social network itself. Instead, some are also capable of using physical opportunities provided the device (e.g., blocking app updates, screen time management). This example illustrates how the existence of these generic physical opportunities provided by the operating system can influence the privacy management capability of young TikTok users to learn about additional ways to protect their privacy. 

In line with previous research we found that negative past experiences affect future privacy management behaviors \cite{livingstoneChildrenDataPrivacy2019}. Incidents can even serve as a learning opportunity \cite{wisniewskiPrivacyParadoxAdolescent2018}. In our sample, participants experienced near or actual privacy incidents (e.g., accidentally publishing videos, loss of account with personal videos) that led them to adapt their privacy management (e.g., immediately deleting accidentally published videos, paying more attention to a publication in the future). While our data support the hypothesis that incidents serve as learning opportunities, it must be said that certain very extreme violations of privacy (e.g., persistent bullying or stalking) have not been reported in our study. It is unclear how such experiences affect privacy behavior in the long run. Nonetheless, our findings inform future research by showing that even minor privacy incidents without severe consequences can lead to improved capabilities.

\subsection{Physical Opportunities}
Adolescents in our sample used various features of TikTok and the operating system to manage their privacy (themes platform features and device features). At the same time, they were aware of TikTok's privacy management limitations (e.g., the ineffectiveness of blocking users). Some of the measures TikTok has taken to protect the privacy of younger users in response to public criticism may not be very effective. Out of 29 study participants with whom we discussed the topic, two-thirds used a false age. Many teenagers we interviewed have been publishing on TikTok much before the legally allowed age of 13. Regardless of the normative standpoint, this calls into question TikTok's fine-grained, age-based privacy features. Despite legislative measures such as the Children's Online Privacy Protection Act of 1998, this problem has been described on other social networks in the past \cite{livingstone2011social, livingstoneRisksSafetyInternet2011}. Sometimes also parents help their underage children to access social networks \cite{boydWhyParentsHelp2011}. Reasons for using social networks below the specified minimum age are diverse (e.g., wanting to stay in touch with classmates, wanting unrestricted access to TikTok's features) \cite{boydWhyParentsHelp2011}. Consequently, technical measures to protect children such as non-public accounts or content restrictions are failing \cite{oneillWhoCaresPractical2013}. boyd et. al \cite{boydWhyParentsHelp2011} called for abandoning ineffective age-based mechanisms. Instead, she advocates for an honest discussion about children's use of social media and a rethinking of the industry to better incorporate the needs of children and parents when developing apps. 

Another issue on social networking sites is account loss \cite{ rathoreSocialNetworkSecurity2017}. This issue was also highlighted by several of our respondents who reported that they were unable to reclaim a video they had posted after losing an account. As a consequence, they were unable to revoke their consent from publishing a childhood experiment that would now remain online forever. This is particularly problematic against the background of increasingly better algorithms for recognizing people in images and videos and the resulting linkability risk (e.g., Clearview AI \cite{hillSecretiveCompanyThat2020}). To exercise the "right to be forgotten" as embodied in the EU GDPR, for example, the ability to reclaim accounts and delete old videos is essential. It is unclear whether account loss among adolescents is a broader phenomenon or whether other social networks are affected as well.

\subsection{Social Opportunities}
Our findings on TikTok support previous research demonstrating that the social environment of teenagers shapes their privacy behaviors \cite{livingstoneChildrenDataPrivacy2019}. Other social network users as well as the parents are major agents of socialization \cite{fengTeensConcernPrivacy2014}. Social norms, which emerge as a response to observed behavior or expected attitudes of friends and parents, influence children's intention to share personal information \cite{vangoolShareNotShare2015}. If friends and parents disapprove of such behavior, children tend to share less. A recent study on TikTok described, that restrictive mediation by parents can also lead to more restrictive disclosure behavior in children \cite{kangTeensPrivacyManagement2021}.

In our study, we identified similar social influences on TikTok. Observing strangers being publicly criticized for videos (theme negative feedback) resulted in restrictive publication behavior by the adolescents we interviewed. In line with previous research \cite{vangoolShareNotShare2015}, the restrictive norms and behavior of relatives, parents, and friends were also found to have the potential to affect behavior on TikTok (e.g., not publishing or blocking parents from videos).

What makes TikTok stand out from other social networks, is its specific content algorithm based on a granular observation of user preferences \cite{klugTrickPleaseMixedMethod2021}. The results of our study indicate that prevalent TikTok usage among peers in combination with the platform's specific algorithm that immediately displays the published content to cohorts with similar attributes – i.e., peers – may increase the social influence of others on adolescents' privacy behavior (“linkability experience”). Unlike posting a video under a nickname on YouTube that may never be discovered by peers, adolescents were aware that posting on TikTok was potentially more privacy-invasive. They recognized that their videos could become visible to their personal environment (e.g., in the schoolyard). This experience led to restricted publication behavior.

\subsection{Automatic and Reflective Motivations}
Adolescents' motivations for protecting their privacy on TikTok were based on either wanting to avoid publicity, to avoid negative reactions/emotions, or to actively achieve privacy (themes negative reaction avoidance, publicity avoidance). The adolescents interviewed reported wanting to evade the public eye and feared negative feedback (e.g., public criticism). These are themes previously described on other social networks \cite{livingstoneChildrenDataPrivacy2019}. To avoid a negative emotional outcome (e.g., shame), they refrain from having a too public profile (theme negative emotion avoidance) (see \cite{buchiChillingEffectsDigital2022} for a similar finding). 

For some adolescents, privacy was a personal matter beyond TikTok (theme privacy identity). That is, these teenagers were intrinsically motivated to keep their information private - a finding that stands in contrast with previous research on other social networks. Research suggests that, on average, adolescents have fewer privacy concerns than young adults \cite{barnesPrivacyParadoxSocial2006, dhirOnlinePrivacyConcerns2017}. However, our findings indicate that these concerns can vary greatly across adolescents, and some may place great value on their privacy on social media. Even though the theme was mentioned by only a few participants, it underscores that adolescents are not a homogeneous group when it comes to motives for managing privacy on social media. For some participants' being private is a personal value and their goal is to achieve a coherent privacy behavior on TikTok and beyond. 

\subsection{Methodological Consideration}
For our study, the COM-B model helped to holistically understand adolescents' privacy management on TikTok related to the creation of videos. It has a solid theoretical foundation and – according to its authors – can be applied across various contexts. However, much of the research to date has applied the COM-B model to health-related behaviors such as smoking cessation and lowering cardiovascular disease risk \cite{michieBehaviourChangeWheel2014}. Our study, which showed that the COM-B model is also a suitable analytical framework for studying privacy behavior, provides yet another use case. By demonstrating its relevance to the privacy management of adolescents, we strengthen the model's extrinsic validity. 

\subsection{Possible Approaches for Privacy Interventions}
\label{section:intervention}
\begin{figure*}[t]
 \centering
 \includegraphics[width=1.4\columnwidth]{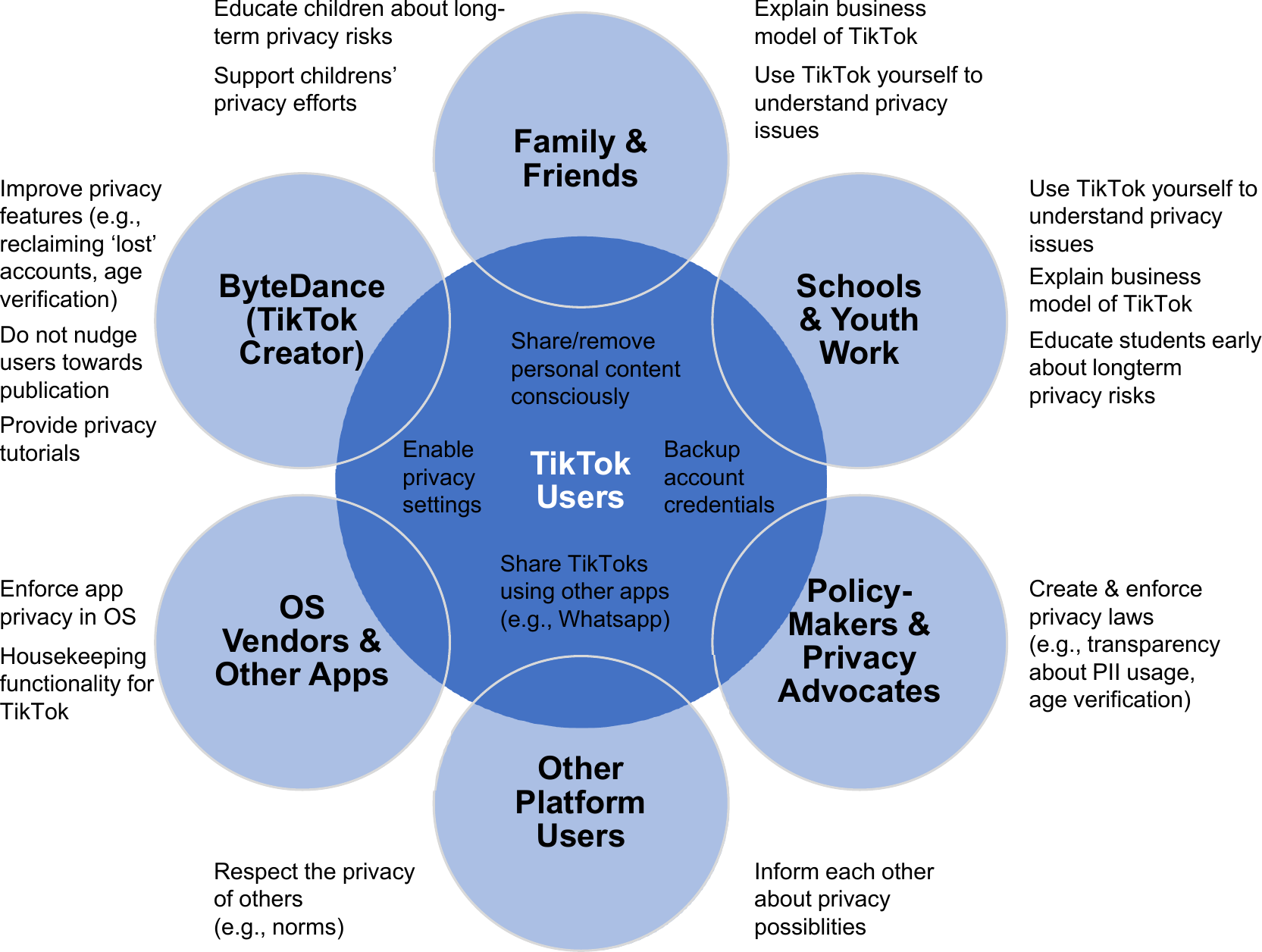}
 \Description{The picture shows a middle circle (TikTok Users) surrounded by six smaller circles (Family & Friends, Schools & Youth Work, Policy Makers & Privacy Advocates, Other Platform Users, OS Vendors & Other Apps, ByteDance). On each of the circles, exemplary behaviors are noted that could be influenced by interventions.}
 \caption{Different parties and their potential target behaviors relevant for adolescents' video privacy management on TikTok}
 \label{fig:circles}
\end{figure*}

Several of the themes we identified can be used as starting points for the development of privacy interventions. The COM-B model is part of a theory-driven intervention development framework called behavior change wheel (BCW), a synthesis of behavior change frameworks \cite{michieBehaviourChangeWheel2011}. In the logic of the BCW, interventions are directed at desired “target behaviors” (e.g., enabling privacy settings). Building on the interview findings and our observations, Figure \ref{fig:circles} shows different parties and ideas for potential target behaviors affecting adolescents' video privacy management. It focuses on \textit{which} behaviors to address and does not answer the question of \textit{how} to design interventions that address these behaviors (e.g., adequate behavior change techniques \cite{michieBehaviorChangeTechnique2013}). 

Any intervention schemes to improve the privacy of adolescent TikTok users should focus on the \textit{behavior of the adolescents themselves}. The interviews provide concrete suggestions for behaviors that adolescents already report that improve their privacy protection. This includes encouraging young users to remove inappropriate videos from the platform and the use of alternative social media apps (e.g., WhatsApp) to share content (theme: proactive privacy). Some of our participants reported regular checks if a video with the status “published” should be set to private. They also removed their old TikToks from the app and their smartphone. Our private creators did seldom publish on TikTok but used alternative apps such as Snapchat or WhatsApp with a perceived higher level of privacy and the ability to automatically delete shared TikToks after being watched by their friends. Another possible target behavior derived from our observations is “backing up user credentials” (theme: privacy literacy). Some adolescents in our sample who had already created accounts in musical.ly could not delete published videos because they had forgotten their credentials, and were not able to prove their identity to the TikTok support to retrieve their account. An intervention could mitigate account loss, especially in cases where children have multiple accounts. Finally, teenagers should be made aware of the privacy settings (e.g., the private account) and the potential risks of not correctly setting these (theme: reflective motivation). For example, in our interviews, participants accidentally published a TikTok upon their first usage of the app because they were not aware others would immediately see it.

\textit{The platform} must also play an important role in safeguarding the privacy of children and adolescents. Improving features directly related to privacy such as improved age verification, more effective blocking of users, and facilitating access to lost user accounts are promising approaches (theme: platform features). As described earlier, many adolescents in our sample did not use their real age due to various reasons. For example, they were often unaware that the private account would have been activated by default if they had provided their real age. As a result of providing false information, the privacy settings were much more lenient and TikTok videos would not only be published to followers but to everybody. Following boyd et. al's \cite{boydWhyParentsHelp2011} philosophy, one possibility would be to abandon TikTok's age-based mechanism and incorporate the needs of children and parents when developing the app. For TikTok, this could mean taking a certain level of responsibility for its content and giving kids and parents ways to control what videos are shown (e.g., via a content configuration or a separate app similar to YouTube Kids). Even if the app adhered to the age-based privacy concept, describing the consequences of providing real age (e.g., better privacy protection) might encourage some youth to provide their real age. Another approach has been lately launched by the twin app Douyin \cite{fengChineseVersionTikTok2021}. Douyin introduced an age verification that is not based on self-declaration only but requires – unlike the international counterpart TikTok - user authentication and imposes restrictions on the permitted daily use for users under 14.

Some participants also criticized that they could not effectively block users who they wanted to prevent from seeing their videos. The problem persists because blocked users can immediately "respawn" under a different username. TikTok could prevent this issue with a feature that block all accounts of the same user (similar to Instagram \cite{WhatHappensWhen}). Some study participants also reported feeling “nudged” by the user interface design towards publishing TikTok video for a broad audience. Others described publishing personal TikToks accidentally. While nudging teenagers towards better privacy behavior is also controversial \cite{veretilnykovaNudgingChildrenAdolescents2021}, presenting them with simple alternatives (such as publishing a TikTok vs saving a local draft) could provide a welcome middle ground. Furthermore, TikTok might also do more to educate its users on how to protect their privacy. This suggestion is based on our observation that capabilities varied between adolescents and TikTok users had begun to create such privacy tutorials. The latter indicates a demand for more support (e.g., via privacy tutorials provided by TikTok). 

\textit{Family, friends, schools, and youth workers} can also positively influence the privacy management of adolescents (social opportunity themes). In addition to supporting adolescents' privacy efforts, their social network could use TikTok themselves to better understand specific privacy issues. In our sample, an uncle of an eight-year-old boy used TikTok himself and warned him about the possibility on TikTok of publishing a video by accident. The social environment can also advise about long-term privacy risks to the children and adolescents of which they might not yet be aware. Among a group of adolescents of the same class, we repeatedly heard the narrative of a classmate being recognized on TikTok despite her wearing a mask. Due to this “risk narrative” the whole class was aware of the potential risks of insufficient anonymization on TikTok. A collection of such tales could be used by teachers in the classroom to illustrate the privacy risk associated with the platform. 

As users do not only interact with each other when they share videos but also with the platform and its owner company, teenagers should also be made aware of commercial privacy issues. Our data confirmed that adolescents' primary privacy focus was indeed social. To this end, adolescents would need to understand TikTok's business model, which heavily relies on their personal data, and the organization behind TikTok.

\textit{Policymakers and privacy advocates} are also relevant actors. Not only do they seek to create privacy laws to protect users but also to enforce these laws through, for example, insisting on effective age verification (theme: platform features). Ideally, these actions are guided by evidence in collaboration with researchers, adolescent users, and parents. For example, our findings indicate that adolescents did not know that TikTok had taken additional measures to protect them in 2021 \cite{terripetersHowTikTokNew2021}. While privacy legislation demands transparency for data subjects – especially for children – this example shows that there is room for improvement in terms of the implementation of laws. 

It should also be mentioned that \textit{other TikTok users} can influence an adolescent's privacy behavior (social opportunity themes). Older and more experienced teenagers may have capabilities (e.g., based on their negative experiences) that can benefit younger and less experienced users. One of our participants reported having learned about privacy settings from a video on TikTok. Indeed, some more experienced users have already begun to acts as mentors. This includes the user @seansvv with 1.1 million followers, who stated in his biography “I Read ToS [Terms of Service] So That You Don't Have To” and regularly posts TikTok videos related to privacy topics \cite{sean[they/them]seansvvTikTok}.

Finally, our interviews showed that \textit{OS vendors and the vendors of other apps} contribute to teenagers' privacy on TikTok (theme: device features). OS vendors have implemented more and more privacy control mechanisms for their end-users (e.g., granular rights management, location sharing notifications). These methods all work on low-level personal data (e.g., IP address, location, and email address). However, videos shared by adolescents on TikTok that possibly contain more sensitive personal data with higher risks involved are not yet covered by these mechanisms. At times when a user publishes a video accidentally, the OS could warn them in the same way that they are warned when sharing their location with the app. In our sample, participants reported also manually cleaning up their TikToks in the app and on their phones. OS vendors could provide housekeeping functionalities that would simplify removing personal content across different social networks and on the phone. 

\subsection{Limitations and Future Research}
As with most qualitative research, our sample is small and was not drawn randomly. Therefore, we cannot claim that the results are representative of all young people in the region under consideration, and certainly not of Switzerland as a whole. Further validation with different samples is needed to strengthen the findings (e.g., including subjects' socioeconomic status).

Choosing interviews as our data collection methodology was useful to learn more about the perspectives of adolescents in Switzerland. Nevertheless, we are aware of the limitations associated with this method. Primarily, we relied on self-reporting rather than behavioral observations. Self-reports can be biased due to various influences, such as subjects' desire to portray themselves in a positive light. Future studies might want to gather data from a wider range of sources, such as direct observations of privacy management behavior (e.g., through TikTok data donations).

Based on our findings, future research could develop and systematically test privacy interventions based on the BCW methodology. A necessary first step would be to identify appropriate target behaviors with the greatest potential to improve privacy management among adolescents. Our research could be a starting point for select a “promising” target behavior reported by the adolescents (e.g., activating the private account) to address in a target population (e.g., pupils of a local school). To identify a baseline for each of the potential behaviors and to select a target behavior among them, further research would be necessary (e.g., in form of a survey among pupils). Furthermore, additional research is required to select appropriate behavior change techniques (e.g., increasing awareness for privacy settings) and evaluate their effectiveness (e.g., with an experiment). Importantly, such research could also control for factors such as socioeconomic status might also be relevant to explain privacy-related behaviors on TikTok \cite{hargittaiDigitalNaIves2010}. Given that teenagers may have very heterogeneous privacy management capabilities, motivations, and opportunities, depending on their age and experience regarding the platform, interventions need to be tailored to the specific target group. Large-scale intervention studies using the BCW can help to identify effective and evidence-based policies to improve privacy management among young people on social media platforms like TikTok.

Our interviews focused on social aspects of adolescents' privacy management. That is, our interviewees were more concerned with protecting their privacy from their social environment than from the corporations dealing with their data for commercial purposes; see \cite{livingstoneChildrenDataPrivacy2019}. Yet, TikTok videos are not only shared with other users but also with ByteDance. Even the users we identified as pure consumers who only view but not create content may have privacy issues. As the video and ad algorithms are known for their high level of customization, they make the platform heavily reliant on personal data including detailed user behavior \cite{klugTrickPleaseMixedMethod2021}. Both users' active and passive behavior on the app has consequences: The TikTok pixel allows companies to engage in detailed web tracking of TikTok users on websites (e.g., a user who sees the ad on TikTok might buy the product in the online shop) \cite{bytedanceltd.InstallTikTokPixel}. Further research could investigate if adolescent users are aware of these commercial privacy aspects and how they manage them. 

\begin{acks}
 The research reported in this article was funded by the Digital Future Fund (DFF), which is part of the Digitalization Initiative of the Zurich Higher Education Institutions (DIZH), Switzerland. We would like to thank all adolescents, teachers, and social workers we contacted in conducting our study. We also thank Frank Wieber, Katja Kurz and Manuel Günther for their helpful comments.
\end{acks}


\bibliographystyle{ACM-Reference-Format}
\bibliography{pets}

\appendix
\section{Appendix}
\subsection{Interview Guide (translated from German)}
\begin{enumerate}
\item What's your first name? How old are you? In what grade are you?
\item How often do you use TikTok? How long have you been using TikTok? When did you start to use TikTok?
\item Do you remember how old you were when you started using the app?
\item How many people do you follow? How many followers do you have?
\item Do you share videos? How many? What types of videos? (Behavior)
\begin{enumerate}
\item Why do you/don't you share videos? (Motivation)
\item If yes: How do you share videos on TikTok? (Psychological capability)
\end{enumerate}
\item Who can see your videos when they are shared? (Psychological capability)
\item How can you influence who can see your videos? (Psychological capability)
\item Do you restrict who can see your videos? (Behavior)
\begin{enumerate}
\item If yes: Why / When do you restrict your videos? (Motivation)
\item If no: Why? Have you ever considered restricting your videos? (Motivation)
\end{enumerate}
\item Do your friends or others restrict their/your videos? (Social Opportunity)
\item Have you ever accidentally posted a video? If yes: What did you do? (Psychological capability)
\item What do you think about TikTok's features to share/restrict videos? (Physical opportunity)
\end{enumerate}

\subsection{Coding Scheme}
Table \ref{tab:codingscheme} shows the hierarchical coding scheme together with the frequency of each code calculated across the 54 interviews.
\begin{table*}[b!]

 \caption{Coding Scheme (translated from German). Frequency is calculated across 54 interviews.}
 \label{tab:codingscheme}

 \begin{tabular}{lp{10cm}c}
\hline
Code&Description&Frequency\\
\hline
Usage\_since&	Start of TikTok usage&	54\\
Usage\_frequency&	Frequency of TikTok usage&	53\\
App\_age&	Age entered into the app at first use&	29\\ [0.5ex]
Video\_Behavior&	&	\\
\hspace{2mm}Avoidance&	I normally do not create/publish TikToks. &	24\\
\hspace{2mm}Proactive&	&	\\
\hspace{4mm}PersonalCreator&	I regularly create/publish TikToks for myself and close friends. &	19\\
\hspace{4mm}PublicCreator&	I regularly create/publish TikToks for my followers/the public. &	11\\ [0.5ex]
Video\_PsyCapability&	&	\\ 
\hspace{2mm}PastPrivacyIncidents&	&	\\
\hspace{4mm}Minor&	I have perceived a potential/minor privacy incident. &	9\\
\hspace{4mm}Severe&	I have perceived a severe privacy incident. &	8\\
\hspace{2mm}PrivacyLiteracy&	&	\\
\hspace{4mm}AudienceContentLiteracy&	I'm aware of different audience/content types and have the ability to manage them. &	52\\
\hspace{4mm}TechnicalLiteracy&
I have the technical knowledge and skills to manage my audience. &	50\\ [0.5ex]
Video\_SocOpportunity&	&	\\
\hspace{2mm}NegativeFeedback&	Others show negative reactions to TikToks, that's why I'm not active. &	16\\
\hspace{2mm}LinkabilityExperience&Users can be easily recognized in real life. & 39\\
\hspace{2mm}RestrictiveInfluence&	I'm not active because others are also restrictive or enforce my privacy. &	34\\ [0.5ex]
Video\_PhyOpportunity&	&	\\
\hspace{2mm}PlatformFeatures&	TikTok helps to ensure my privacy. &	46\\
\hspace{2mm}DeviceFeatures&	The device helps to ensure my privacy. &	17\\ [0.5ex]
Video\_AutMotivation&	&	\\
\hspace{2mm}NegativeEmotionAvoidance&	I don't publish content to avoid negative emotions. &	15\\ [0.5ex]
Video\_RefMotivation&	&	\\
\hspace{2mm}NegativeReactionAvoidance&	I don't publish content to avoid negative reactions. &	10\\
\hspace{2mm}PrivacyIdentity&	I don't publish content because privacy is important to me. &	5\\
\hspace{2mm}PublicityAvoidance&	I don't publish content because I don't want to be in the public eye. &	29\\
\hline
\end{tabular} 

\end{table*}
\end{document}